\documentclass[letterpaper, 10 pt, conference]{ieeeconf}
\IEEEoverridecommandlockouts                         
\usepackage{amsmath}
\usepackage{amssymb}
\usepackage{amsthm}
\newtheoremstyle{mystyle}%
  {\topsep}%
  {\topsep}%
  {\normalfont}%
  {}%
  {\bfseries}%
  {}%
  {.5em}%
  {}%
\theoremstyle{plain}

\newcommand{\norm}[1]{\left\lVert#1\right\rVert}
\usepackage{mathtools}
\usepackage{siunitx}

\usepackage{algorithm}
\usepackage{algpseudocode}

\usepackage{lipsum}

\usepackage{subcaption}
\usepackage{subfig}

\usepackage[short,nocomma]{optidef}

\usepackage{circuitikz}

\title{\LARGE \bf
Safe Learning-Based Optimization of Model Predictive Control: Application to Battery Fast-Charging
}

\author{Sebastian Hirt$^{1}$, Andreas Höhl$^{1}$, Johannes Pohlodek$^{1}$, Joachim Schaeffer$^{1, 2}$, Maik Pfefferkorn$^{1}$,\\ Richard D. Braatz$^{2}$ and Rolf Findeisen$^{1}$%
\thanks{$^{1}$Technical University of Darmstadt, Germany {\tt\small \{sebastian.hirt, rolf.findeisen\}@iat.tu-darmstadt.de}, $^{2}$Massachusetts Institute of Technology, MA, USA}%
}

\usepackage{tikz}

\newcommand\submittedtext{%
  \footnotesize This work has been submitted to the IEEE for possible publication. Copyright may be transferred without notice, after which this version may no longer be accessible.}

\newcommand\submittednotice{%
\begin{tikzpicture}[remember picture,overlay]
\node[anchor=south,yshift=10pt] at (current page.south) {\fbox{\parbox{\dimexpr0.65\textwidth-\fboxsep-\fboxrule\relax}{\submittedtext}}};
\end{tikzpicture}%
}

\begin{document}

\maketitle
\submittednotice
\thispagestyle{empty}
\pagestyle{empty}

\begin{abstract}
Model predictive control (MPC) is a powerful tool for controlling complex nonlinear systems under constraints, but often struggles with model uncertainties and the design of suitable cost functions. To address these challenges, we discuss an approach that integrates MPC with safe Bayesian optimization to optimize long-term closed-loop performance despite significant model-plant mismatches.
By parameterizing the MPC stage cost function using a radial basis function network, we employ Bayesian optimization as a multi-episode learning strategy to tune the controller without relying on precise system models.
This method mitigates conservativeness introduced by overly cautious soft constraints in the MPC cost function and provides probabilistic safety guarantees during learning, ensuring that safety-critical constraints are met with high probability.
As a practical application, we apply our approach to fast charging of lithium-ion batteries, a challenging task due to the complicated battery dynamics and strict safety requirements, subject to the requirement to be implementable in real time.
Simulation results demonstrate that, in the context of model-plant mismatch, our method reduces charging times compared to traditional MPC methods while maintaining safety.
This work extends previous research by emphasizing closed-loop constraint satisfaction and offers a promising solution for enhancing performance in systems where model uncertainties and safety are critical concerns.
\end{abstract}

\section{Introduction}
Model predictive control (MPC) is a powerful technique for optimal control of complex, nonlinear systems, particularly when constraints on system states and control inputs are present \cite{findeisen2002introduction,rawlings2017model}.
Despite its strengths, MPC faces significant challenges due to its reliance on accurate system models and appropriately designed cost functions.
Obtaining accurate models is often difficult because of the complexity of the system and limited insights into the process.
In addition, selecting a suitable cost function that yields the desired closed-loop performance while ensuring safety remains a challenging task.

Machine learning has emerged as a promising solution to address these issues by enabling data-driven model learning within the MPC framework \cite{hewing2020cautious,maiworm2021online,zieger2020towards}.
However, incorporating machine learning into control systems raises concerns about performance, stability, and repeated feasibility \cite{Mesbah2022}.
Although robust MPC formulations can tackle these challenges, they tend to be computationally expensive, can result in overly conservative control policies, and rely on uncertainty bounds that are difficult to obtain for models based on machine learning.
Moreover, even highly accurate prediction models do not always guarantee optimal closed-loop performance with respect to a higher-level performance metric \cite{kordabad2023reinforcement}.

Recent work has explored hierarchical frameworks that combine optimization-based controllers with learning algorithms to overcome these limitations.
These frameworks typically involve a two-layer structure comprised of a high-level layer, which is focused on long-term and global performance optimization and safety, and a low-level predictive controller, which is responsible for short-term planning and control.
The latter often uses a prediction horizon significantly shorter than the duration of the task.
The high-level layer often uses techniques from reinforcement learning \cite{kordabad2023reinforcement,zanon2021safe} or Bayesian optimization \cite{paulson2023tutorial,piga2019performance,hirt2024learning}, working in conjunction with the lower-level controller.
This hierarchical approach offers a promising means to achieve the desired closed-loop behavior while addressing performance limitations caused by insufficient process knowledge.

In this work, we address the optimization of long-term closed-loop performance for nonlinear dynamical systems exhibiting significant model-plant mismatch due to limited process insights, specifically focusing on fast-charging of lithium-ion batteries.
We are motivated by the need to develop control methods for safe fast charging that do not increase battery degradation or compromise safety in the presence of a model-plant mismatch, while being applicable in real time \cite{krewer2018,kim2024fast}.
By adhering to safety constraints, the battery lifetime is prolonged, which is important for users and, in particular, for the adoption of electric vehicles.

To this end, we employ a multi-episode learning approach utilizing Bayesian optimization \cite{garnett2023bayesian}, a sample-efficient black-box optimization method, to tune the parameters of a model predictive controller with short prediction horizons, thus extending the results from \cite{matschek2023, hirt2024learning}.
By parameterizing the MPC stage cost function using a radial basis function network, we propose an alternative to traditional model learning, taking advantage of the intrinsic connection between model-based predictions and the MPC cost function.
Simultaneously, our approach provides a direct solution to mitigate the conservativeness introduced by overly cautious soft-constrained penalty terms in the MPC cost function.

This work extends our previous research on cost function learning \cite{hirt2024safe, hirt2024stability}, where we focused on neural network-based parameterizations and closed-loop stability during the learning process.
In contrast, the present work emphasizes closed-loop constraint satisfaction and radial base function-based parameterization.
The advantage of using radial basis functions is that they only act locally in the state space where they are needed, making them ideal for selectively relaxing soft constraint penalty terms in the cost function and offering better interpretability compared to the weights and biases of neural networks.
Additionally, we generalize our previous approaches towards considering a distribution of initial states in the closed-loop experiments and apply the approach towards safe fast charging of lithium-ion batteries.

The remainder of this work is structured as follows. Section \ref{sec:fundamentals} provides an introduction to parameterized MPC, Gaussian process regression for surrogate modeling, and safe Bayesian optimization.
Afterward, the safe learning approach is introduced in Section \ref{sec:main}, including the radial-basis-function-based parameterization of the MPC and the incorporation of information about closed-loop constraint satisfaction into the safe Bayesian optimization procedure.
Section \ref{sec:simulation} describes the battery simulation case study and results that highlight the performance and safety capabilities of our framework.
We draw conclusions in Section \ref{sec:conclusion}.

\section{Fundamentals}
\label{sec:fundamentals}
In this section, we define the control objective and introduce a parameterized model predictive control.
Next, we present Gaussian process regression for surrogate modeling, followed by an overview of safe Bayesian optimization.

\subsection{Problem Formulation}
We focus on nonlinear, discrete-time dynamical system
\begin{equation}
\label{eqn:discrete_system_general}
\begin{split}
    x_{k+1} &= f(x_k, u_k), \\
    y_k &= h(x_k, u_k),
\end{split}
\end{equation}
where $x_k \in \mathbb{R}^{n_\text{x}}$ are the system states, $u_k \in \mathbb{R}^{n_\text{u}}$ are the system inputs, $y_k \in \mathbb{R}^{n_\text{y}}$ are the system outputs, $f: \mathbb{R}^{n_\text{x}} \times \mathbb{R}^{n_\text{u}} \to \mathbb{R}^{n_\text{x}}$ is the nonlinear dynamics, $h: \mathbb{R}^{n_\text{x}} \times \mathbb{R}^{n_\text{u}} \to \mathbb{R}^{n_\text{y}}$ is the output mapping, and $k \in \mathbb{N}_0$ is the discrete time index.

Our control objective is to optimally steer the system \eqref{eqn:discrete_system_general} from an initial state $x_0$ to a desired set point $(x_d, u_d)$, while satisfying all input constraints $\forall k: u_k \in \mathcal{U}$ and state constraints $\forall k: x_k \in \mathcal{X}$.

 We assume a significant mismatch between the controller's internal prediction model and the actual plant dynamics, leading to suboptimal closed-loop performance and challenges with closed-loop constraint satisfaction.
Additionally, we only use short prediction horizons to reduce computational complexity and enable real-time implementation, which limits the controller's ability to anticipate future system behavior.
To address these challenges, we parameterize the MPC formulation.
This adds degrees of freedom to the controller, allowing us to adjust aspects like the controller's cost function to better compensate for the model-plant mismatch and enhance closed-loop performance, even with limited prediction horizons.
Consequently, our primary goal becomes to find the appropriate parameters that optimize closed-loop performance while ensuring closed-loop constraint satisfaction.
To learn such a parameterization in a structured and safe manner, we employ safe Bayesian optimization based on closed-loop data.

\subsection{Parameterized Model Predictive Control}
We consider an MPC formulation that includes $n_{\text{p}} \in \mathbb{N}$ parameters $\theta \in \Theta \subset \mathbb{R}^{n_{\text{p}}}$.
Thus, at each discrete time step $k$, given a specific parameter set $\theta$, the MPC solves a parameterized optimal control problem defined by
\begin{mini!}
    {\mathbf{\hat{u}}_k}{\left\{\sum_{i=0}^{N-1} l_\theta({\hat x}_{i \mid k}, {\hat u}_{i \mid k}, {\hat y}_{i \mid k}) + E_\theta({\hat x}_{N \mid k}) \! \right\}\label{eqn:mpc_ocp_cost}}{\label{eqn:mpc_ocp}}{}
    \addConstraint{\forall i}{\in \{0, 1, \dots, N-1\}: \notag}{}
    \addConstraint{}{\hat x_{i+1\mid k} = \hat f_\theta(\hat x_{i\mid k}, \hat u_{i\mid k}), \ \hat x_{0 \mid k} = x_k,}{\label{eqn:mpc_ocp_model}}
    \addConstraint{}{\hat y_{i\mid k} = \hat h_\theta(\hat x_{i\mid k}, \hat u_{i\mid k})},{\label{eqn:mpc_ocp_model_output}}
    \addConstraint{}{\hat x_{i \mid k} \in \mathcal{X}_\theta, \ {\hat u}_{i \mid k} \in \mathcal{U}, \ \hat x_{N \mid k} \in \mathcal{E}_\theta.}{\label{eqn:mpc_ocp_constraints}}
\end{mini!}
where $\hat{\cdot}_{i\mid k}$ denotes the model-based $i$-step ahead prediction at time index $k$, $\hat f_\theta$ and $\hat h_\theta$ are the (parametric) prediction and output models, respectively, and $x_k$ is a measurement of the true system state at time index $k$.
We consider a finite prediction horizon of length $N \in \mathbb{N}$, and the (parametric) stage and terminal cost functions $l_\theta$ and $E_\theta$.
The constraints \eqref{eqn:mpc_ocp_constraints} consist of the (parametric) state, input, and terminal sets $\mathcal{X}_\theta$, $\mathcal{U}$, and $\mathcal{E}_\theta$, respectively.
Solving \eqref{eqn:mpc_ocp} results in the optimal input sequence $\mathbf{\hat{u}}_k^*(x_k; \theta)=[\hat u_{0 \mid k}^*(x_k; \theta),\dots,\hat u_{N-1 \mid k}^*(x_k; \theta)]$, and we apply its first element to system \eqref{eqn:discrete_system_general}.

\subsection{Gaussian Process Surrogate Models}
\label{sec:gp}
To learn the parameters of the MPC \eqref{eqn:mpc_ocp} using Bayesian optimization from closed-loop data, it is necessary to construct a surrogate model that captures the relationship between the parameters $\theta$ and the closed-loop performance, as well as the potential closed-loop constraints.
For this purpose, we apply Gaussian process (GP) regression to the closed-loop data generated by the system \eqref{eqn:discrete_system_general} when controlled by MPC \eqref{eqn:mpc_ocp}.

We employ Gaussian processes to model an unknown function $\varphi: \mathbb{R}^{n_\xi} \to \mathbb{R}, \xi \mapsto \varphi(\xi)$ probabilistically from finitely many observations.
Loosely speaking, a GP $g(\xi) \sim \mathcal{GP}(m(\xi), k(\xi, \xi^\prime))$ is a generalization of the Gaussian probability distribution to an infinite-dimensional function space.
A GP is fully defined by its prior mean function $m: \mathbb{R}^{n_\xi} \to \mathbb{R}, \xi \mapsto \mathrm{E}[g(\xi)]$ and its prior covariance function $k: \mathbb{R}^{n_\xi} \times \mathbb{R}^{n_\xi} \to \mathbb{R}, (\xi, \xi') \mapsto \mathrm{Cov}[g(\xi), g(\xi')]$.

Our objective is to develop a predictive model for $\varphi$ that can accurately estimate the unobserved function values $\varphi(\xi_*)$ at test inputs $\xi_*$.
To achieve this, we use a set of (noisy) function observations $\mathcal{D} = \left\{ (\xi_i, \gamma_i = \varphi(\xi_i) + \varepsilon_i) \mid i \in {1, \dots, n_{\text{d}} }, \varepsilon_i \sim \mathcal{N}(0, \sigma^2) \right\}$, where $\varepsilon_i$ represents Gaussian white noise with variance $\sigma^2$.
We collect all training inputs $\xi_i$ in the input matrix $\Xi \in \mathbb{R}^{n_{\text{d}} \times n_\xi}$ and the associated training targets in the vector $\gamma \in \mathbb{R}^{n_{\text{d}} \times 1}$, so that the data set can be equivalently represented as $\mathcal{D} = (\Xi, \gamma)$.
Bayesian inference is then employed to incorporate the information provided by the training data into the model.

The Bayesian inference step yields the posterior distribution $g(\xi_*) \mid \Xi, \gamma, \xi_* \sim \mathcal{N}(m^+(\xi_*), k^+(\xi_*))$ with mean and variance given by
\begin{subequations}
\label{eqn:posterior_gp}
\begin{align}
    m^+(\xi_*) &= m(\xi_*)+k(\xi_*,  \Xi) k_{\gamma}^{-1} (\gamma-m(\Xi)), \label{eq:gp_postMean} \\
    k^+(\xi_*) &= k(\xi_*, \xi_*) - k(\xi_*, \Xi) k_{\gamma}^{-1} k(\Xi, \xi_*). \label{eq:gp_postVar}
\end{align}
\end{subequations}
where $k_{\gamma} = k(\Xi, \Xi) + \sigma^2 I$ and $I$ is the identity matrix.
The posterior mean \eqref{eq:gp_postMean} is the best estimate of the unknown function value $\varphi(\xi_*)$, and the posterior variance \eqref{eq:gp_postVar} quantifies the uncertainty of the predicted value \cite{rasmussen2006gaussian}.

The prior mean and covariance functions are design choices that typically involve several hyperparameters. These hyperparameters must be adjusted to the underlying problem to ensure a good fit to the training data.
A common approach to determining the appropriate hyperparameters is to infer them from the available training data through evidence maximization \cite{rasmussen2006gaussian}.

\subsection{Safe Bayesian Optimization}
Bayesian optimization (BO) is an iterative approach designed for efficient optimization of expensive-to-evaluate black-box functions \cite{garnett2023bayesian}.
In this work, we employ BO to improve the closed-loop performance of the system \eqref{eqn:discrete_system_general} controlled by MPC \eqref{eqn:mpc_ocp}, which depends on the controller parameters $\theta$, while ensuring compliance with predefined safety constraints.
Since the functional relationships between the parameters $\theta$ and the closed-loop performance and safety measures are complex, non-analytical, and expensive to evaluate, BO is particularly well-suited for this task.

Formally, we employ BO to solve the optimization 
\begin{align}
\begin{split}
    \label{eqn:bo_optimization_problem}
    \theta^* &= \arg \max_{\theta \in \Theta} \left\{G_0(\theta)\right\} \\
    & \text{s.t.} \  G_i(\theta) \geq 0, \ \forall i \in \{1, \ldots, n_{\text{bc}} \},
\end{split}
\end{align}
where $G_0$ is the closed-loop performance and the $G_i$ are $n_{\text{bc}} \in \mathbb{N}$ black-box (safety) constraints.
Since $G_0$ and $G_i$ are unknown or expensive to evaluate, BO relies on probabilistic surrogate models derived from observations.
We employ Gaussian processes as surrogate models, which is a common choice because of their efficiency when working with limited training data.
Throughout the learning process, the GP surrogate models are iteratively updated with data collected from exploring different MPC parameterizations, thereby establishing the sequential nature of the BO procedure.
In each iteration $n \in \mathbb{N}$, we first select the next set of parameters of interest $\theta_n$, and perform a closed-loop run using MPC \eqref{eqn:mpc_ocp} to generate a new data point $\{ \theta_n, G_0(\theta_n), \ldots, G_{n_{\text{bc}}}(\theta_n) \}$.
Next, we update the training data set by incorporating this newly observed data point $\mathcal{D}_{n+1} \leftarrow \mathcal{D}_n \cup \{ \theta_n, G_0(\theta_n), \ldots, G_{n_{\text{bc}}}(\theta_n) \}$, where $\mathcal{D}_n$ denotes the training data set with data points observed up to iteration $n$.
Finally, we update the posterior GP models based on $\mathcal{D}_{n+1}$, which includes optimizing the hyperparameters.

To systematically perform the sequential learning procedure, an acquisition function is used to guide the selection of new parameter sets $\theta_{n}$ toward the optimal parameters $\theta^*$ as $n$ increases. 
The acquisition function $\alpha_0: \mathbb{R}^{n_{\text{p}}} \to \mathbb{R}, \theta \mapsto \alpha_0(\theta; \mathcal{D}_n)$ leverages the surrogate model of the unknown function $G_0$ to assess the utility of a parameter set $\theta_n$ for improving closed-loop performance.
Using the uncertainty information from the surrogate model, BO effectively balances exploration and exploitation when selecting the parameters for each closed-loop run.

We determine the next set of parameters according to
\begin{equation}
    \label{eqn:bo_update}
    \theta_{n+1} = \arg \max_{\theta \in \Theta} \left\{ \alpha_0(\theta; \mathcal{D}_n) + \tau \ \sum_{i=1}^{n_\text{bc}} \alpha_i(\theta; \mathcal{D}_n) \right\}.
\end{equation}
Here, the term $\alpha_0(\theta; \mathcal{D}_n)$ is focused on closed-loop performance, while $\alpha_i: \mathbb{R}^{n_{\text{p}}} \to \mathbb{R}, \theta \mapsto \alpha_i(\theta; \mathcal{D}_n)$ is used to include black-box constraints $G_i, i \in \{1, \dots, n_\text{bc} \}$, and $\tau \in \mathbb{R}_+$ is the penalty parameter.
As penalty terms, we use logarithmic barriers given by
\begin{equation}
    \label{eqn:log_barrier_term}
    \alpha_i(\theta;\mathcal{D}_n) = \log \!\left(m_{G_i}^+(\theta) - \beta(\delta) \sqrt{k_{G_i}^+(\theta)} \right),
\end{equation}
where $m_{G_i}^+$ and $k_{G_i}^+$ are the posterior mean and variance of the GP model of $G_i$ and $\beta$ is the confidence scaling parameter, depending on the probability of constraint satisfaction given by $1-\delta$.
By incorporating the lower confidence bound in the selection of the next set of parameters to evaluate, it is possible to probabilistically guarantee fulfillment of the posed black-box constraints, given that the GP models are well-calibrated, i.e., the confidence intervals are exact.
See \cite{hirt2024safe} and \cite{krishnamoorthy2023tuning} for a more detailed discussion of the theoretical guarantees.

\section{Safe Learning of an RBF-Network-based Cost Function}
\label{sec:main}
This section describes our approach to achieving safe closed-loop learning.
We introduce a radial basis function parameterization of the MPC stage cost function and detail a safe Bayesian optimization procedure to effectively learn its parameters.

\subsection{Cost Function Parameterization}
In our simulation scenario, we start with a safe but conservative MPC.
Conservativeness is introduced by overly cautious soft constraint penalties in the MPC cost function, motivated by the posed safety requirements and the assumption of a significant but unknown model-plant mismatch.
To counteract conservativeness, we introduce a parameterization in the stage cost using a radial basis function (RBF) network.
Specifically, we choose the MPC stage cost as
\begin{equation}
    l_\theta (x, u) = l(x, u) + l_{\text{RBF}}(x, u; \theta),
\end{equation}
where $l$ consists of the known and potentially conservative objective and the parameterized component is given by
\begin{align}
    l_{\text{RBF}} (x, u; \theta) &= \sum_{i=1}^{n_{\text{RBF}}} w_i \rho(\norm{x-c_i}).
\end{align}
Herein, $n_{\text{RBF}} \in \mathbb{N}$ is the number of radial basis functions, $w_i \in \mathbb{R}$ and $c_i \in \mathbb{R}^{n_{\text{x}}}$ are the weights and centers for each RBF, respectively, and $\rho$ is the radial basis function, e.g., a Gaussian given by
\begin{equation}
    \label{eqn:gaussian_rbf}
    \rho(\norm{x-c_i}) = \exp(-\lambda_i \norm{x-c_i}^2).
\end{equation}
To limit the number of parameters that must be learned, we choose the centers $c_i$ and widths $\lambda_i \in \mathbb{R}$ of the RBF to be fixed and only parameterize the weights $w_i$.
This results in the parameterization $\theta = \{ w_1, w_2, \dots, w_{n_{\text{RBF}}} \}$.
In general, the centers and widths can also be parameterized.

\subsection{Closed-loop Constraint Satisfaction during Learning}
Deriving safety guarantees for closed-loop state constraint satisfaction, is challenging due to the varying parameterization of the MPC.
This is especially the case for the parameterization chosen via an RBF network, which provides a high degree of freedom.
To ensure safe operation, we enforce state constraints (probabilistically) via constraints in the BO optimization.
We assume box constraints on the states given by
\begin{equation}
    \mathcal{X} = \left\{ x \in \mathbb{R}_{n_\text{x}} \;\big|\; x_{\text{min}} \leq x \leq x_{\text{max}} \right\},
\end{equation}
where the inequality is understood component-wise and $x_{\text{min}} = \left[ x_{\text{min}}^1, \ldots, x_{\text{min}}^{n_\text{x}} \right]^\intercal$, $x_{\text{max}} = \left[ x_{\text{max}}^1, \ldots, x_{\text{max}}^{n_\text{x}} \right]^\intercal$.
Based on this requirement, we define the BO constraints for closed-loop constraint satisfaction as
\begin{align}
    \label{eqn:closed_loop_constraint}
    G_i(\theta) = \min_{k \in \{0, \cdots{}, M\}} \left\{ x_{\text{max}}^i - x_k^i \right\},
\end{align}
where $x_k^i$ is the $i$th component of the state vector at sampling time $k$, i.e., $x_{k} = \left[ x_{k}^1, \ldots, x_{k}^{n_\text{x}} \right]^\intercal$ and $M \in \mathbb{N}$ is the length of the closed-loop trajectory.
\eqref{eqn:closed_loop_constraint} penalizes the largest constraint violation occurring during a closed-loop run with parameterization $\theta$ for the $i$th state component.
In this way, all critical safety states are taken into account during the learning procedure.
The constraint formulation \eqref{eqn:closed_loop_constraint} enforces compliance of the component of the $i$th state $x_k^i$ with its upper limit $x_{\text{max}}^i$.
An individual constraint is analogously formulated for all state components, as well as for the lower bound.
The right-hand side of \eqref{eqn:closed_loop_constraint} implicitly depends on the controller parameters through the closed-loop state trajectory.
By choosing $\beta(\delta)$, satisfaction of the closed-loop constraints during learning is guaranteed with a probability of $1-\delta$ \cite{hirt2024safe}.

\section{Simulation Study}
\label{sec:simulation}
This section demonstrates the improvements of the safe learning framework for a battery fast-charging case study. 

\subsection{Battery Fast-charging}
Lithium-ion batteries (LIBs) exhibit time-varying, nonlinear dynamics, which can be modeled using several approaches.
These include empirical methods such as lookup tables, black-box machine learning techniques such as neural networks, first-principles models such as porous electrode theory models, and equivalent circuit models (ECMs) \cite{schaeffer2024}.
ECMs approximate battery dynamics through an electric circuit composed of a voltage source, resistors, capacitors, and other elements \cite{krewer2018,schaeffer2023}.
Unlike first-principles models, such as the Doyle-Fuller-Newman model, ECMs offer a lower computational cost and a straightforward implementation, making them popular for embedded control applications.

\begin{figure}[!t]
    \vspace{2mm}
    \centering
    \scalebox{.99}{
    \begin{circuitikz}%
        \draw (0,0) to [voltage source, v=${\text{OCV} (z_k)}$] (0,-2.4);
        \draw (0,0) to [resistor, l=$R_0 (z_k)$] (3,0)
                    to [resistor, l=$R_1 (z_k)$] (5,0);
        \draw (3,0) -- (3,-1.3)
                    to [C, l=$C_1 (z_k)$] (5,-1.3)
                    -- (5,0) to[short, i=$I$, -*] (6.3,0);
        \draw (0,-2.4) to[short, -*] (6.3,-2.4);
        \draw (6.3,0) to[open, v=$V_T$] (6.3,-2.4);
    \end{circuitikz}
    }
    \caption{R-RC battery ECM with parameters depending on the SOC $z_k$.}
    \label{fig:ecm}
\end{figure}
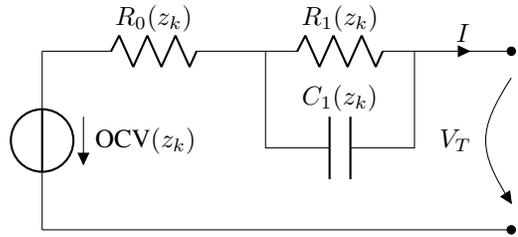
We use an ECM consisting of a resistor and a resistor in series with a resistor-capacitance pair for both the plant and the MPC model.
However, we introduce a significant model-plant mismatch of up to 50\% by disturbing the parameters of the prediction model.
The following fast-charging case studies utilize an ECM consisting of a resistor in series with a resistor-capacitance pair (Fig.~\ref{fig:ecm}). 
The open circuit voltage (OCV) depends on the state of charge (SOC) and defines the ideal voltage source.
The battery's ohmic resistance is represented by $R_0$, while the polarization caused by the double layer and the diffusion are approximated by the $R_1$-$C_1$ pair.

The battery's state $x_k = [z_k, U_{1, k}, T_k]$ is comprised of the battery's SOC $z_k$, the voltage over the capacitor $U_{1, k}$, and its temperature $T_k$.
The control input is the charging current $u_k = I_k$ and the output is the terminal voltage $y_k = V_{T, k}$.
The battery dynamics are given by
\renewcommand{\arraystretch}{1.4}
\begin{equation} 
\label{eqn:discrete_system}
\begin{split}
    \begin{bmatrix}
        z_{k+1} \\
        U_{1, k+1} \\
        T_{k+1}
    \end{bmatrix} &\!=\!
    \begin{bmatrix}
        z_k + \frac{\eta \Delta t}{Q} I_k\\
        (U_{1,k}-R_1 I_k)\exp\!{\big(\!\!-\!\frac{\Delta t}{R_1 C_1}\big)} \!+ R_1 I_k\\
        T_k+\frac{\Delta t}{C_\text{th}}\big(I_k^2(R_0+R_1)-\frac{T_k-T_\text{amb}}{R_\text{th}}\big)
    \end{bmatrix}\!, \\
    V_{T, k} &= \mathrm{OCV} - U_{1,k} - R_0 I_k,
\end{split}
\end{equation}
with the SOC-dependent circuit elements ($R_0$, $R_1$, $C_1$, $\mathrm{OCV}$), the sampling time $\Delta t$, the Coulombic efficiency $\eta$, and the battery discharge capacity $Q$.
The difference equation for the change in temperature additionally depends on the thermal capacity $C_\text{th}$, the thermal resistance $R_\text{th}$, and the ambient temperature $T_\text{amb}$.

We used cubic splines to interpolate the published parameters for a Samsung SDI INR18650-20S lithium nickel manganese cobalt oxide (NMC) cell \cite{tran2021}.
For simplicity, we neglect the current and temperature dependencies of the electric circuit elements. 
Model-plant mismatches are introduced in practice for the following reasons: (1) The model is always only an approximation of the battery, (2) parameters are usually interpolated and uncertain, (3) battery degradation with time and current throughput alters system dynamics and reduces capacity and power capabilities.

The fast charging of LIBs is important for improving the user experience of electric vehicles and other consumer devices.
An approach to define fast charging protocols is to minimize charging time while meeting degradation constraints \cite{matschek2023}. 
Typically, these constraints focus on current, voltage, temperature, and lithium-plating overpotential to limit degradation.
In this work, we implement constraints on current, voltage, and temperature given by
\begin{align}
    \forall k \in \mathbb{N}_0&:\; I_k \leq I_{\mathrm{max}} = \SI{6}{\ampere} \label{eq:imin},\\ 
    \forall k \in \mathbb{N}_0&:\;\SI{2.5}{\volt}=V_{T,\mathrm{min}} \leq \;V_{T,k} \leq V_{T,\mathrm{max}} = \SI{4.2}{\volt}\label{eq:vtmin},\\
    \forall k \in \mathbb{N}_0&:\; T_k\leq T_\text{max}=\SI{318}{\kelvin}.\label{eq:tmin}
\end{align}
The plating overpotential constraint cannot be applied because the selected ECM does not model plating.
Due to the model-based approach, we maintain current control throughout the charging process, rather than switching to voltage control after reaching the upper cutoff voltage \cite{chen2006}.

\subsection{Unconstrained Bayesian Optimization for Closed-loop Learning}
In this first simulation study, we consider unconstrained BO as a baseline.
To account for varying initial conditions, for example, battery temperature, we randomly sample the initial condition $T_0$ for each closed-loop run.
We define the closed-loop objective as the mean of the cumulative tracking error over a set of trajectories, each initialized with different initial conditions, given by
\begin{equation}
    \label{eqn:battery_closed_loop_performance}
    G_0(\theta) = \frac{1}{n_{\text{i}}} \sum_{l=1}^{n_{\text{i}}} \sum_{k=0}^M (1-z_k^l)^2
\end{equation}
where $z_k^l$ is the state of charge (SOC) at the $k$th sampling time for trajectory $l$, 
the index $l$ corresponds to the initial condition $l \in \{ 1, \cdots{}, n_{\text{i}} \}$, and $n_{\text{i}} \in \mathbb{N}$ denotes the number of distinct initial conditions considered in each Bayesian optimization iteration.
The index $k$ runs from 0 to $M \in \mathbb{N}$, where $M$ is the total number of time steps in each trajectory.

We choose the known part of the MPC stage cost function to encode our objective of fully charging the battery and fulfill the safety critical constraints on the terminal voltage and the temperature.
The employed stage cost is given by
\begin{align*}
    \label{eqn:battery_stage_cost_analytic}
    l(x_k) & = (1-z_k)^2 \\
    & + \gamma_T \cdot \max\left( 0, T_k - (T_{\text{max}} - \Delta T_{\text{max}}) \right)^2 \\
    & + \gamma_{V_T} \cdot \max\left( 0, V_{T_k} - (V_{T, {\text{max}}} - \Delta V_{T, \text{max}}) \right)^2,
\end{align*}
where $\gamma_T$ and $\gamma_{V_T}$ are penalty weights, $T_{\text{max}}$ and $V_{T, {\text{max}}}$ are the maximum allowable temperature and terminal voltage, respectively, and $\Delta T_{\text{max}}$ and $\Delta V_{T, \text{max}}$ are slack variables.
We choose the slack variables such that the constraints are fulfilled in the closed-loop.
However, this approach introduces conservativeness into the closed-loop performance of the MPC formulation.
This is practically motivated by the fact that the slack variables $\Delta V_T$ and $\Delta T$ are not known in the case of a significant and unknown model-plant mismatch.
Thus, they are chosen conservatively in this initial MPC formulation.
To counteract the conservativeness introduced by the penalty terms, we additionally parameterize the MPC stage cost function.
Specifically, we choose RBFs as in \eqref{eqn:gaussian_rbf} with fixed centers $c_i$, fixed widths $\lambda_i$, and parameterized weights.
To not compromise our objective of charging the battery, the RBF cost only depends on $V_{T, k}$ and $T_k$ and is given by
\begin{align}
    l_{\text{RBF}} (V_{T, k}, T_k; \theta) &= \sum_{i=1}^{n_{\text{RBF}}} w_i \rho(\norm{[V_{T, k}, T_k]-c_i}),
\end{align}
where $c_i \in \mathbb{R}^2$.
We choose 16 RBFs and thus have to learn their weight parameters via the BO procedure.
In this first simulation study, we choose not to formulate constraints in the BO optimization problem to establish a baseline on the performance of the learning and control framework.

\begin{figure}
    \centering
    \vspace{3mm}
    \includegraphics[width=0.48\textwidth]{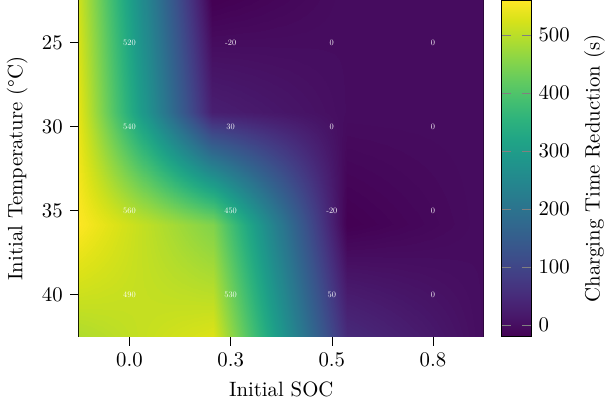}
    \caption{Reduction in charging time to 80\% SOC using unconstrained BO.}
    \label{fig:charging_time_advantage_unconstrained_bo}
\end{figure}

The results using unconstrained BO are shown in Fig. \ref{fig:charging_time_advantage_unconstrained_bo}.
We illustrate the impact of initial state of charge (SOC) and initial temperature on the charging time to 80\% SOC.
The figure shows that for some initial conditions, particularly low initial SOC values and moderate temperatures around 30°C to 35°C, the unconstrained BO approach leads to significant reductions in charging time, with improvements of up to 560 seconds over the initial unparameterized MPC.
However, in other regions, such as at lower temperatures and higher initial SOC values, the optimization results in little to no reduction, and in some cases even negative values, indicating that the charging time has increased relative to the unparameterized MPC.
Furthermore, the constraints \eqref{eq:vtmin} and \eqref{eq:tmin} were violated in 16 of 40 BO iterations, indicating unsafe learning.
Consequently, in the following simulation study, we consider the incorporation of closed-loop constraints into the BO procedure.

\subsection{Safe Bayesian Optimization for Closed-loop Learning}
For safe learning, we extend the setup from the previous section using constraints defined by \eqref{eqn:closed_loop_constraint} and account for the worst-case violation of constraints during the $n_\text{i}$ runs observed in each BO iteration.
We specifically consider the safety critical constraints on the terminal voltage and the temperature given in \eqref{eq:vtmin} and \eqref{eq:tmin}.
In our closed-loop constraint formulation, we choose the confidence scaling parameter $\beta = 1$, see \eqref{eqn:log_barrier_term}, providing a suitable trade-off between probabilistic constraint satisfaction and performance.

Since the non-parametric MPC formulation is conservative but safe, we choose $\theta = 0$ as our initial safe parameterization, that is, all RBF weights are equal to zero.
Now, the objective is to choose parameters that optimize closed-loop performance \eqref{eqn:battery_closed_loop_performance} while not violating constraints during learning.

\begin{figure}
    \centering
    \vspace{3mm}
    \includegraphics[width=0.48\textwidth]{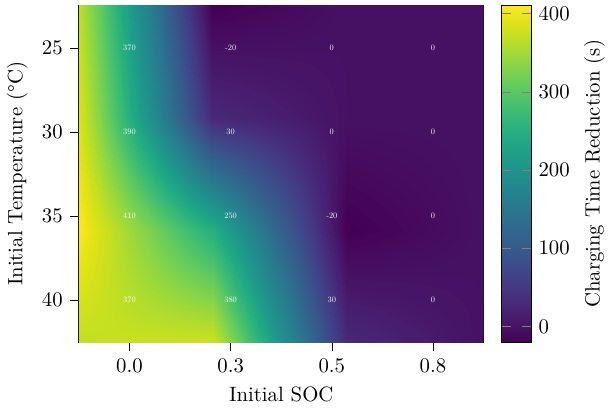}
    \caption{Reduction in charging time to 80\% SOC using safe BO.}
    \label{fig:charging_time_advantage_safe_bo}
\end{figure}

\begin{figure}
    \vspace{3mm}
    \hspace{\fill}
    \begin{subfigure}{0.6\textwidth}
        \includegraphics{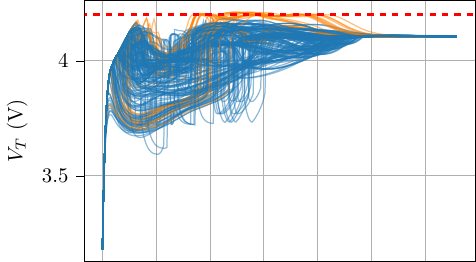}
    \end{subfigure}
    \begin{subfigure}{0.6\textwidth}
        \includegraphics{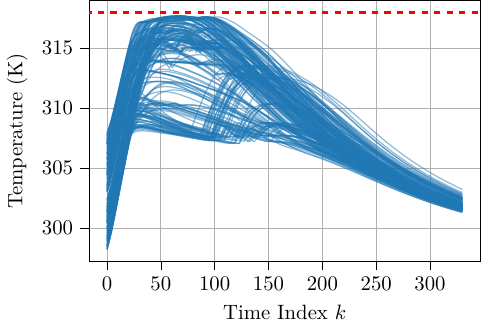}
    \end{subfigure}
    \vspace{-3mm}
    \caption{Terminal voltage $V_T$ (top) and temperature $T$ (bottom) trajectories for all sampled charging cycles during the safe learning procedure, using a confidence scaling parameter of $\beta = 1$. Constraints are shown in dashed red and unsafe trajectories are shown in orange.}
    \label{fig:constraints_safe_bo}
\end{figure}

The results are shown in Fig. \ref{fig:charging_time_advantage_safe_bo}.
Compared to the unconstrained BO approach, safe BO results in slightly lower reductions in charging time, with the best performance observed at low initial SOC values and moderate temperatures, reaching a reduction of up to 410 seconds.
Again, the reductions remain positive or zero in a wider range of initial conditions, highlighting significantly improved performance.
Importantly, closed-loop constraints were violated in only 6 of 40 safe BO iterations, compared to 16 of 40 violations with the unconstrained approach, indicating a substantial improvement in safety while still achieving competitive performance; see Fig. \ref{fig:constraints_safe_bo}.
Our closed-loop constraint ensures probabilistic safety and, thus, allows for some constraint violations.
However, the ratio of $6/40 = 0.15$ constraint violations using safe BO is well below the maximum allowed violation rate of $0.32$ defined by the confidence scaling parameter $\beta = 1$ (one standard deviation), while the ration $16/40 = 0.4$ for unconstrained BO was above the maximum allowed rate.
Furthermore, the maximum constraint violation for $V_T$ was $6.6\%$ lower in amplitude using safe BO compared to unconstrained BO.
For $T$, no constraint violation occurred.
Note that the probability of closed-loop constraint satisfaction given by $1-\delta$ can be effectively adjusted using the confidence scaling parameter $\beta(\delta)$ in \eqref{eqn:log_barrier_term} \cite{hirt2024safe}.

\section{Conclusions}
\label{sec:conclusion}
We showcased a learning and control methodology that combines safe Bayesian optimization with model predictive control (MPC) to address the challenge of optimizing fast-charging controllers for lithium-ion batteries.
By parameterizing the MPC cost function using a radial basis function network and leveraging a safe learning framework, our approach enables cost function learning while ensuring probabilistic closed-loop constraint satisfaction.
This allows the controller to adapt its behavior based on closed-loop data, mitigating conservativeness without sacrificing safety.
Our simulation results for a lithium-ion battery case study show that the proposed method significantly reduces charging times compared to traditional MPC formulations, while providing probabilistic safety guarantees during the learning process.
This work extends prior research by integrating constraint satisfaction directly into the cost function learning procedure, offering a solution for optimization of long-term closed-loop performance in safety-critical applications.

\bibliographystyle{ieeetr}
\bibliography{bibliography}

\end{document}